\documentclass[prd]{revtex4}
\usepackage{array}
\usepackage{booktabs}
\usepackage{tabu}
\usepackage{dcolumn}
\usepackage{amsmath}
\usepackage{amsfonts}
\usepackage{amssymb}
\usepackage{graphicx}
\usepackage{subfigure}
\usepackage{graphicx}
\usepackage{dcolumn}
\usepackage{bm}

\begin{document}

\title{Stability of  a noncanonical  scalar field model during cosmological date}
  \author{Z. Ossoulian$^{a}$}
   \email{zossoulian@gmail.com}
    \author{T. Golanbari$^{a}$}
    \email{t.golanbari@gmail.com/ @uok.ac.ir}
    \author{H. Sheikhahmadi$^{a,b}$}
    \email{h.sh.ahmadi@iasbs.ac.ir/ gmail.com}
      \author{Kh. Saaidi$^{a}$}
       \email{ksaaidi@uok.ac.ir}
\affiliation{$^a$Department of Physics, Faculty of Science, University of Kurdistan,  Sanandaj, Iran.\\
$^b$Institute for Advance Studies in Basic Sciences (IASBS), Gava Zang, Zanjan 45137-66731, Iran.}
\date{\today}

\def\be{\begin{equation}}
  \def\ee{\end{equation}}
\def\bea{\begin{eqnarray}}
\def\eea{\end{eqnarray}}
\def\f{\frac}
\def\n{\nonumber}
\def\l{\label}
\def\p{\phi}
\def\o{\over}
\def\R{\rho}
\def\pa{\partial}
\def\om{\omega}
\def\na{\nabla}
\def\P{\Phi}
\def\g{\gamma}

\begin{abstract}
\section*{Abstract}
Using the noncanonical model of scalar field,   the cosmological consequences of a pervasive, self-interacting, homogeneous and  rolling  scalar field  are studied.  In this model, the scalar field potential is "nonlinear" and decreases in magnitude with  increasing  the value of the scalar field. A special solution of the nonlinear field equations of $\phi$  that has  time dependency as fixed point is obtained. The fixed point relies on the noncanonical term of action and $\gamma$-parameter, this parameter is appeared in energy density of scalar field red shift. By means of such fixed point the different eigenvalues of the equation of motion will be obtained. In different epochs in the evolution of the Universe for different values of $q$ and $n$ the potentials as a function of scalar field are attained. The behavior of baryonic perturbations in linear perturbation scenario as a considerable  amount of energy density of  scalar field at low red shifts prevents the growth of perturbations in the ordinary matter  fluid. The energy density in the scalar field is not appreciably perturbed by non-relativistic gravitational fields,  either in the radiation or matter dominant, or scalar field dominated epoch.
\end{abstract}
\pacs{.....}
\maketitle
\section{Introduction}\label{Intro}
During two past decades, undoubtedly people believe the Universe is undergoing an accelerated expansion phase. This interesting and unanticipated result comes from some observations including Cosmic Microwave Background  (CMB) \cite{CMB_th_1, CMB2}, Supernovae type Ia (SNeIa) \cite{ch1:7, ch1:8}, Baryonic Acoustic Oscillations (BAO) \cite{BAO_th_Eisenstein, BAO_th_Shoji}, Observational Hubble Data (OHD) \cite{FarooqRatra, H0_1}, Sloan Digital Sky Survey (SDSS) \cite{ch1:10, SD2}, and Wilkinson Microwave Anisotropy Probe (WMAP)~\cite{ch1:9,WM2}. To justify such accelerated epoch, people mostly consider an unknown form of matter which produces a negative pressure namely dark energy\cite{Bamba}. These observations also anticipate  that the nonluminous   components of the universe (dark enery $68.3\%$, dark matter $26.6\%$ \cite{Ade01, Ade02}) are $94.9\% $ and ordinary  matter (baryonic an radiation) is only $5.1\%$.  It is notable that study the origin and nature of dark energy the best candidate is cosmological constant, c.c., $\Lambda$ \cite{3, 3a, 3d,  3b, 3e}. But unfortunately this interesting attractive candidate of dark energy suffers two  well-known problems, fine tuning and coincidence problems. The former moot point  refers to the difference between the theoretical anticipation of the energy density of $\Lambda$ and what is risen from observations. The latter comes back to this question, why the ratio of matter and dark energy densities in present epoch is so close to unit?\cite{4}. These two problems convince scientists to make other proposals to investigate the  behaviour of dark energy. Accordingly people are interested to search for some proposals and models which resolve two mentioned problems and also have a strong bankroll from both theoretical and experimental point of views. People unusually has proposed different proposals based on two completely different perspectives, where the one of them is a geometrical scenario and the other is related to the manipulation the matter components of the Universe. For the former model one can mention $f(R)$ {\cite{fr1, fr2, fr3, fr4}}, $f(T)$ {\cite{ft1, ft2, ft3, ft4, ft5}}, $f(G)$ {\cite{fg1, fg2, fg3, fg4}} gravities and modification or combination of them based on cosmological purpose, and for latter for example we can name some models which are risen from quantum gravity (e.g. holographic model) {\cite{13O, 13aO, 13bO, 13cO, 13dO}},  space time fluctuations (e.g. age and new-age graphic models) {\cite{12O, 14O, 14aO, 15O}}, vacuum quantum fluctuations (e.g. Casimir effect in large scale) {\cite{12O, 11O, 11aO, My}} and also we are able to allusion the scalar field proposals.  It is well known that scalar field scenario usually is considered as the first and best candidate for new phenomena which were appeared in physics. For instance Yukawa model of strong force {\cite{Yu, Yu1}}, Landau-Ginzburg mechanism {\cite{L-G}}, Higgs mechanism {\cite{H, H1, H2}}, inflation \cite{I, I1, I2}, \cite{Ia, Ia1} and so on. Also it is notable whereas scalar fields under the general coordinate re-parametrization have the simplest behaviour, naturally has attracted more attention recently. It is considerable as a special categorize, we can split the scalar field scenarios to canonical and noncanonical models. For the canonical proposal one can mention Brans-Dicke {\cite{brans1, brans2}}, quintessence {\cite{5, 5a, 5c, 5b}}, phantom {\cite{8, 8-1, 8a, 8b}}, quintom {\cite{9, 9a, 9b}}, and chameleon {\cite{11}, \cite{11a}, \cite{10}, \cite{ch18}, \cite{ch19} and \cite{ch20, ch20-1, ch1:20a, ch1:18}}. Also for noncanonical mechanism we can refer  k-essence {\cite{6, 6a, 6b}}, Dirac-Born-Infeld model {\cite{DBI}}, tachyon {\cite{7}} and so on. It should be stressed, the noncanonical mechanism has some advantages against canonical mechanism. For instance although quintessence proposal as a tracker model re-solves coincidence problem, it needs to be highly tuned; It is notable in noncanonical scenario both coincidence and fine tuning problems could be re-solved. Another interesting aspect of the noncanonical model (k-essence) against quintessence is that, the kinetic term in such scenario can source the c.c. These interesting aspects of noncanonical model motivated us to investigate the stability of this scenario in different epoches in cosmology. For this aim  we are actually interested in the behavior, in linear perturbation theory, of density perturbations in the scalar field energy density and in the densities of baryons and radiation. It is obvious that scalar field energy density fluctuations tend to decay inside of the horizon, while in a baryon dominated universe the behavior of baryonic perturbations are effectually the same as in the canonical scenario. On the other hand, if the energy density in the homogeneous part of the scalar field is considerable, ordinary matter perturbations cannot grow. Since scalar field perturbations do not grow inside the horizon, the scalar-field energy density will remain very much smoother than the baryonic, so the peaks in the matter distribution will be almost entirely baryonic. The mass of the scalar field fluctuation (second derivative of the scalar potential) is related to the horizon size; the scalar-field fluctuation is extremely light. If the scalar field were to dominate early enough, it would suppress growth of baryonic structure on small scales. This is because the Universe expands so fast that the perturbations cannot collapse. Note that the scalar field has to be minimally coupled to matter or must be very weakly coupled to ordinary matter so that it does not drag the matter perturbations with it and thereby prevent them from collapsing, even before the scalar field comes to dominate the energy density of the Universe. This is one reason why we have to suppose  that the Universe has only recently become dominated by the scalar-field energy density, if galaxies formed by gravitational instability. This effect motivates the choice of potentials $V(\phi)$ discussed below. We study  a class model of the  potential of the scalar field. In this  model, we assume that the energy density of the scalar field red shifts in a certain way and then determines the potential of the scalar field that is required. We find out  that the scalar field potential is "nonlinear" which  typically  tends to be made from negative powers of the scalar field.  We have not been succeeded in determining the general solution of the scalar field equation of motion, but a special solution can
be found (in which the scalar field energy density red shifts in the requisite manner). Somewhat remarkably, we find that this solution dominates at large time and a study of phase space shows that it is an attractive and time dependent fixed point (in the cases of interest, it is the only attractive fixed point in phase space). This solution may, therefore, be chosen as a background solution for a study of the evolution of density inhomogeneities in linear perturbation theory.\\
The scheme of this paper is as follows, in Sec.$I$ some aspects of both canonical and non-canonical mechanisms, and also, the importance of scalar field proposal beside the motivations of this work were discussed as introduction. Sec.$II$  the general framework of this work includes of the math calculations and also the equations of energy density and pressure of noncanonical model will discussed, also we will obtain the general form of the potential for this model.  In Sec.$III$ by means of a perturbation mechanism, we will solve the equation of motion of scalar field. To solve this equation we should find out the fix point of the model, and by using it the eigenvalues of the equation will be obtained, based on this values we will discuss different conditions to investigate the stability or non-stability of the model.  In Sec.$IV$ we will consider different epoches in the evolution of the Universe, namely radiation, matter and scalar field dominans respectively.   And at last, results and conclusions of the work will bring in Sec.$V$.

\section{The model}\label{Sec2}
Using canonical model of scalar field the energy density of potential, time derivative of scalar field, the ratio of the energy density of  scalar field to radiation and matter and also the wave equation of scalar field have been studied. In phase space the values of special solution and general solution of wave equation have been obtained and the critical points and perturbations around these critical points in phase space equations have been investigated. Specifying the value of $q$
with respect to the ratio of energy density of scalar field to the radiation and matter,  the dominance of the energy density of scalar field at present time and the positive acceleration of the universe at current epoch have been explained.
In this paper we want to study the above mentioned conditions by means of noncanonical scalar field to get better results in comparison to the results which were obtained in ordinary canonical proposal. It is considerable that, we consider homogeneous, isotropic and spatially flat space time  which is given  by the line element
\begin{equation}\label{metric}
ds^{2}=-dt^{2}+a^{2}( dr^{2}+r^{2}d\theta^{2}+r^{2}\sin^{2}\theta d\phi^{2}) .
\end{equation}
The noncanonical version of Einstein scalar field action is defined as {\cite{6, 6a}}
\begin{equation}\label{action}
\mathcal{S} =\int \sqrt{-g} d^{4}x \left( \dfrac{M_{p}^{2}}{2}R+F(X)-V(\phi)+\mathcal{L}_{m}\right).
\end{equation}
Where $g$ is the determinant of the metric $g_{\mu\nu}$, $R$ is the Ricci scalar, and also $M_p$ is the Planck mass. Potential of scalar field is denoted by $V(\phi)$, $\mathcal{L}_{m}$ is introduced as ordinary matter Lagrangian and the kinetic term of noncanonical scalar field is expressed by an arbitrary function $F(X)$, in which $X=-(g^{\mu\nu}\nabla_{\mu} \phi \nabla_{\nu} \phi)/2$.
It should be stressed that the noncanonical scalar field lagrangian in question  is
\begin{equation}\label{lagSF}
\mathcal{L}_{\phi}=F(X)-V(\phi).
\end{equation}
Variation of  the action (\ref{action}) with respect to $g^{\mu\nu}$ leads to
\begin{equation}\label{Eqfield}
G_{\mu\nu}=R_{\mu\nu}-\frac{1}{2}g_{\mu\nu}R=\dfrac{1}{M_{p}^{2}}\left( T_{\mu\nu}^{m}+T_{\mu\nu}^{\phi}\right),
\end{equation}
By varying the  Eq.(\ref{action}) with respect to $\phi $ and considering Eq.(\ref{metric}), one has
\begin{equation}\label{EqfieldSf}
\big( 2XF_{XX} + F_X \big) \ddot\phi + 3F_X H \dot\phi + \tilde{V}(\phi) = 0,
\end{equation}
where overdot denotes a derivative with respect to cosmic time and $tilde$ indicates derivative with respect to $\phi$.
Here $R_{\mu\nu}$ is the Ricci tensor, $F_{X}= dF(X)/dX$ and $F_{XX}= d^2F(X)/dX^2$. For an arbitrary Lagrangian, $L_{m}$,  the stress-energy tensor is defined as
 \begin{equation}\label{SETensor}
T_{\mu\nu}=\dfrac{-2}{\sqrt{-g}}\dfrac{\delta\left( \sqrt{-g}L_{m}\right)}{\delta g^{\mu\nu}}.
\end{equation}
Therefore stress-energy tensor of scalar field, $T_{\mu\nu}^{\phi}$, is given by
\begin{equation}\label{SETensorSF}
T_{\mu\nu}^{\phi}= g_{\mu\nu}\left[ F(X)-V(\phi)\right] +\left( \nabla_{\mu}\phi\nabla_{\nu}\phi\right) F_{X},
\end{equation}
and $ T_{\mu\nu}^{m}$ is the stress-energy tensor of ordinary matter that for a perfect fluid cosmological matter one hasis given by
\begin{equation}\label{SETensorMatter}
T_{\mu \nu } =(\rho + p)U_\mu U_\nu + pg_{\mu \nu },
\end{equation}
here $ U^\mu $ is four-vector of a fluid. From Eq.(\ref{metric}) we have
\begin{equation}\label{SETensor-diag}
T_{\mu}^{\nu}=diag (-\rho,  p,  p,  p),
\end{equation}
here
$\rho=\rho_{r}+\rho_{m}$ and $p=p_{r}+p_{m}$, where subscripts ${r}$ and ${m}$ are refer to
 energy density of radiation and energy  density of  baryonic and cold dark matter respectively.
 Utilizing the usual definition of the stress-energy tensor, Eqs. (\ref{SETensorSF}) and (\ref{SETensor-diag}), the energy density and the pressure of scalar field are defined as
\begin{subequations}
\begin{eqnarray}\label{DensitySF}
\rho_{\phi} & = & T_{00}^{(\phi)}=2XF_{X}-F(X)+V(\phi),\\\label{PersserSF}
p_\phi & = & T_{ii}^{(\phi)}=F(X)-V(\phi).
\end{eqnarray}
\end{subequations}
By omitting $ \rho_{r} $ and $ \rho_{m} $ and also by means of Eqs.(\ref{Eqfield}), (\ref{SETensorSF}), (\ref{SETensor-diag}) and (\ref{metric}) one can obtain
\begin{equation}\label{FreEq}
H^{2}=\dfrac{8\pi}{3M_{p}^{2}}\left( 2XF_{X}-F(X)+V(\phi)\right) ,
\end{equation}
\begin{equation}\label{AccEq}
{\ddot{a} \over a} = H^2 + \dot{H} = {-1 \over 3M_p^2} \; \Big( F + XF_X - V(\phi) \Big).
\end{equation}
for regarding the spatial  homogeneity of cosmology, we suppose that the scalar field is just a function of time so
 $X=\dot\phi^2 / 2$.
 By considering Eqs. (\ref{DensitySF}) and (\ref{PersserSF}) time derivative of scalar field and the related potential are reduce to
\begin{equation}\label{dotPhi}
\dot\phi^{2}=\dfrac{\rho_{\phi}+p_{\phi}}{F_{X}},
\end{equation}
\begin{equation}\label{VPhi}
V(\phi)=\dfrac{(\rho_{\phi}-p_{\phi})}{2}+F(X)-XF_{X}.
\end{equation}
We suppose, there has not  any interaction between  components of fluid, so that  conservation equation of  stress-energy equation is satisfied for all fluid components of the Universe as
\begin{equation}\label{ConservationEq}
\dot{\rho}_{i}+3H(\rho_{i}+p_{i})=0,
\end{equation}
here $\rho_{i}=\rho_{r}, \rho_{m}, \rho_{\phi}$. In addition, we assume the  scalar field energy  density  red-shifts  is given by
\begin{equation}\label{rhoPhi}
\rho_{\phi}=\rho^{(0)}_{\phi}(\frac{a_{0}}{a})^{q},
\end{equation}
we assume $q\neq0$.
Therefore by means of Eqs. (\ref{ConservationEq}) and (\ref{rhoPhi}) one gets
\begin{equation}\label{PPh}
p_{\phi}=(\dfrac{q-3}{3})\rho_{\phi}.
\end{equation}
where $\omega=(q-3)/3$ is equation of state parameter of scalar field. By considering $F(X)=F_{0}X^{n}$ and using Eqs. (\ref{dotPhi}), (\ref{VPhi}) and (\ref{rhoPhi}) one receives
\begin{equation}\label{dotPhi2}
\dot\phi=\left[ \dfrac{\rho^{(0)}_{\phi}q2^{n-1}}{3F_{0}n}\right]  ^{\frac{1}{2n}}(\frac{a_{0}}{a})^{\frac{q}{2n}}
\end{equation}
\begin{equation}\label{VPhi2}
V(\phi) = \left[ \dfrac{6n - (2n-1)q}{6n} \right] \rho_{\phi}^{(0)}\left( \frac{a_{0}}{a}\right)^{q}.
\end{equation}
It should be noted for $n=1/2$, the coefficient $\ddot{\phi}$ in equation (\ref{EqfieldSf}) is equal to zero, therefore we will not consider this case.
\section{A component  dominated  cosmology}\label{secIII}
As we mentioned in previous sections,  we have assumed that the components of the Universe could be considered as  radiation, matter (ordinary and cold dark matter) and scalar field. But in this  section  we  shall suppose  the contribution of  The universe components in the stress-energy tensor  is only one of them. In fact we assume in any epoch only the energy density  of related component  is dominated and therefore other components of perfect fluid could be  neglected.
According to conservation equation for  energy density, Eq.(\ref{ConservationEq}) and the related equation of state, $p=(\gamma/3-1)\rho$,
the  energy density  of the dominant component  is obtained as
\begin{equation}\label{rhoGamma}
\rho_{\gamma} = \rho_{\gamma}^{(0)}\left( \dfrac{a_{0}}{a} \right)^{\gamma},
\end{equation}
which $ \rho_{\g}^{(0)} $, is the value of energy density at $ a=a_{0} $ and $(\g/3 -1)$ is the equation of state parameter of dominant component. So the Friedman equation (\ref{FreEq}), could be written as follows
\begin{equation}\label{FreEqGamma}
(\frac{\dot{a}}{a})^{2}=\dfrac{\rho_{\g}^{(0)}}{3M_{p}^{2}}\left(\frac{a_{0}}{a}\right)^{\g}.
\end{equation}
Using Eqs. (\ref{dotPhi2}) and (\ref{FreEqGamma}) one obtains
\begin{equation}\label{Phidot-adot}
\dfrac{\dot\phi}{\dot{a}}=\left( \dfrac{q\rho_{\phi}^{(0)}2^{n-1}}{3F_{0}n} \right)^{\frac{1}{2n}}\left( \dfrac{3M_{p}^{2}}{\rho_{\g}^{(0)}a_{0}^{2}}\right) ^{\frac{1}{2}}\left( \frac{a}{a_{0}} \right)^{\frac{\g n-q-2n}{2n}},
\end{equation}
hence by integration the above relation, the scalar field is resulted as
\begin{equation}\label{phi-phi0}
\phi - \phi_{0} = \dfrac{2n}{\g n-q}\left( \dfrac{q\rho_{\phi}^{(0)}2^{n-1}}{3F_{0}n} \right)^{\frac{1}{2n}}\left( \dfrac{3M_{p}^{2}}{\rho_{\g}^{(0)}}\right) ^{\frac{1}{2}}\left( \frac{a}{a_{0}} \right)^{\frac{\g n-q}{2n}},
\end{equation}
where $\phi_{0}$  is  the  value  of  $\phi $   at $a=a_{0}$. And also from  Eqs.(\ref{VPhi2}) and (\ref{phi-phi0}), we have
\begin{equation}\label{V-V0}
V(\phi)  =  V_{0}\left( \phi - \phi_{0} \right)^{\frac{2nq}{q-\g n}},
\end{equation}
where
\begin{widetext}
\begin{equation}\label{V0}
V_{0} =  \dfrac{\left[  6n - (2n-1)q\right]  \rho_{\phi}^{(0)}}{6n}\left[ \dfrac{2n}{\g n-q}\left( \dfrac{q\rho_{\phi}^{(0)}2^{n-1}}{3F_{0}n} \right)^{\frac{1}{2n}}\left( \dfrac{3M_{p}^{2}}{\rho_{\g}^{(0)}}\right) ^{\frac{1}{2}}\right] ^{\frac{2nq}{q-\g n}}.
\end{equation}
\end{widetext}

The  Einstein scalar field  equations (for such potential) have a special  solution. In  this section we study spatially
homogeneous perturbations about this special solution.  To do this we should study the structure of the four-dimensional, spatially homogeneous, phase space ($\phi,\dot\phi$). Therefore, we consider the scalar field equation of motion, which is given by
\begin{widetext}
\begin{equation}\label{y-diffeEq}
y'' + \left( \dfrac{4(n+1)-\g(2n-1)}{2(2n-1)} \right) \dfrac{y'}{x} +  \dfrac{qV_{0}2^{n}}{F_{0}(2n-1)(q-\g n)} \left( \dfrac{3M_{P}^{2}}{\rho_{\g}^{(0)}} \right)^{n}x^{n(\g -2)}\dfrac{ y^{\frac{(2n-1)q+\g n}{q-\g n}}}{\left( y' \right)^{2n-2}} = 0,
\end{equation}
\end{widetext}

where prime indicates  derivative with respect to $x$ and  we have changed the variables as $ x=a/{a_{0}} $ and
$ y=\phi - \phi_{0} $.
A  special  solution of differential equation (\ref{y-diffeEq}) is introduced as
\begin{equation}\label{SpecSol}
y_{e}=kx^{\alpha},
\end{equation}
by substituting Eq.(\ref{SpecSol}) into Eq.(\ref{y-diffeEq}), $k$ and $\alpha$ are obtained as
\begin{subequations}
\begin{eqnarray}\label{C-SpecSol1}
k & = &  \dfrac{2n}{\g n-q}\left[ \dfrac{q\rho_{\phi}^{(0)}2^{n-1}}{3F_{0}n}\right] ^{\frac{1}{2n}}\left( \dfrac{3M_{p}^{2}}{\rho_{\g}^{(0)}}\right)^{\frac{1}{2}},\\\label{C-SpecSol2}
\alpha & = & \dfrac{\g n-q}{2n}.
\end{eqnarray}
\end{subequations}
 To study the structure of the phase space of scalar field's, one can make a change of variable as
\begin{equation}\label{GeneSol}
y(x)=y_{e}(x)u(x).
\end{equation}
In above re-definition $ y_{e}(x) $, the special solution of differential equation, is unperturbed part of scalar field and $ u(x) $ is the perturbed part of the scalar field which should have a stable and attractor property. So for stability of the solutions of scalar field, the differential equation (\ref{y-diffeEq}), should solve for the perturbed part. It is notable that all dynamical information related to equation of motion are at evaluation equation $ u(x) $.
By substituting the general solution into differential equation (\ref{y-diffeEq}) the result is as
\begin{widetext}
\begin{equation}\label{u(x)-diffeEq}
u''+\left[ {2(n+1)\o (2n-1)} -{\g \o 2} +2\alpha\right]  \frac{u'}{x}+\left[ {3\o (2n-1)} -{\g \o 2} +\alpha\right]  \left[ u-\dfrac{u^{\frac{q(2n-1)+\g n}{q-\g n}}}{(\frac{x}{\alpha}u'+u)^{2n-2}}\right]\frac{\alpha}{x^{2}} =0.
\end{equation}
\end{widetext}

By changing  the variable $ x=e^{t} $ and rewriting  Eq.(\ref{u(x)-diffeEq}) in terms of $ (u(t), t) $, one has
 \begin{widetext}
\begin{equation}\label{u(t)-diffeEq}
 \ddot{u}+\left[ {3\o (2n-1)} -{\g \o 2} +2\alpha\right]  \dot{u}+\alpha  \left[ {3\o (2n-1)} -{\g \o 2} +\alpha\right] \left[u-\dfrac{u^{\frac{q(2n-1)+\g n}{q-\g n}}}{(\frac{\dot{u}}{\alpha}+u)^{2n-2}}\right] =0,
\end{equation}
\end{widetext}

 where as former  overdot indicates derivative with respect to $t$. The critical point of this system is defined as a point which the velocities could be neglected. Therefore, the only critical point is obtained as $\dot{u}=0$. Hence according to Eq.(\ref{y-diffeEq}) we have able to study the evolution of $ u(t)$. To do this we use phase space to explain the stability of this nonlinear problem by means of linear analyse. Therefore, in  phase space
$ (u(t), \dot{u}(t)) $, Eq.(\ref{u(t)-diffeEq}) could be rewritten as
\begin{widetext}
\begin{eqnarray}\label{p(t)-diffeEq}
\dot{u}(t)  =& & p(t),\\
\dot{p}(t) = & - & \alpha  \left[ {3\o (2n-1)} -{\g \o 2} +\alpha\right]\left[ u(t)-\dfrac{u(t)^{\frac{(2n-1)q+\g n}{q-\g n}}}{\left( \frac{ p(t)}{\alpha}+u(t)\right)^{2n-2}}\right]  - \left[ {3\o (2n-1)} -{\g \o 2} +2\alpha\right]  p(t) ,\label{dotp(t)-diffeEq}
\end{eqnarray}
\end{widetext}

then for $ p(t)=0 $, we have
\begin{equation}
u(t)^{2n-1} - u(t)^{\frac{q(2n-1)+\g n}{q-\g n}} = 0.
\end{equation}
The point $ (u_{0}, p_{0}) = (1, 0) $ is the only acceptable critical point for the Eqs.(\ref{p(t)-diffeEq}), and (\ref{dotp(t)-diffeEq}).
The system at critical point doesn't has evolution and so the differential equation ,(\ref{y-diffeEq}), just have the special solution, (\ref{SpecSol}). Of course at $ p(t)=0 $, there are a number of critical points, that $u(t)$  is $ \left[(\g n-q)8^{-1}n^{-2}\right] $ the root of unity.
$ u(t) $, should be real, so we don't work with the complex fixed points. It is seen that by changing $ y(x)\rightarrow -y(x) $, the differential equation (\ref{y-diffeEq}), for some values of $ q $ and $ n $, is invariant. In this case there are two critical points, but our purpose is to explain the solutions which $ y'(x)>0 $. So one of the critical points is omitted. The second part of the function $ y(x)= y_{e}(x)u(x) $, $u(x)$ can be perturbed by adding $ u_{1}(t) $ to the critical point $ (u_{0}, p_{0}) $ and substituting at phase space equations (\ref{p(t)-diffeEq}) and (\ref{dotp(t)-diffeEq}). The result is (in fact $ u_{1}(t) $ is  a small fluctuation around the critical point $ (u_{0}, p_{0}) = (1, 0) $)
 added and transform it to the critical point $ \left( u_{0} + u_{1}(t), p_{0}+p_{1}(t)\right)  = \left( 1 + u_{1}(t),  p_{1}(t)\right) $.
\begin{eqnarray}\label{p1(t)-diffeEq}
\dot{u_{1}}(t) & = & 0 + p_{1}(t),\\ \nonumber
\dot{p_{1}}(t) & = &  \g\left[ {q\o 2}-{3n \o 2n-1}\right] u_1(t)+\left[ q-3-{\g \o 2}\right] p_1(t).\label{p2(t)-diffeEq}
\end{eqnarray}
 The eigenvalues of small oscillations around critical point are given by
\begin{eqnarray}\label{eigenvalues}
\lambda_{n, q,\g} = f(n, q,\g) \pm ig(n, q,\g),
\end{eqnarray}
where
\begin{eqnarray}\label{fg-eigenvalues}
f(n, q, \g) & = &\dfrac{2q-6-\g}{4},\nonumber \\
g(n, q, \g) & = & \left[ \g ( {3n\o 2n-1}-{q\o 2}) - {(q-3-{\g\o 2})^2 \o 4}\right] ^{\frac{1}{2}},
\end{eqnarray}
The eigenvalues $\lambda$ for different $n$ and $q$ will be  evaluated.
\section{Investigation of epochs dominant}
In this section we want to examine our results  of previous section for different epochs during evolution of the Universe.

\subsection{Radiation  dominated epoch of cosmology}
We assume the radiation dominant era, and therefore one can neglect the other components of perfect fluid in the Universe. According to conservation equation of radiation, Eq.(\ref{ConservationEq}) and equation of state for radiation, $p=\rho/3$, one can observe that in radiation epoch $\g = 4$. Thus the radiation energy density is resulted as
\begin{equation}\label{4,21}
\rho_{r} = \rho_{r}^{(0)}\left( \dfrac{a_{0}}{a} \right)^{4},
\end{equation}
which $ \rho_{r}^{(0)} $, is the value of radiation energy density at $ a=a_{0} $.
So  according to Eqs. (\ref{phi-phi0}), (\ref{V-V0}) and (\ref{V0}), we have
\begin{equation}\label{24.1}
\phi - \phi_{0} = \dfrac{2n}{4n-q}\left( \dfrac{q\rho_{\phi}^{(0)}2^{n-1}}{3F_{0}n} \right)^{\frac{1}{2n}}\left( \dfrac{3M_{p}^{2}}{\rho_{r}^{(0)}}\right) ^{\frac{1}{2}}\left( \frac{a}{a_{0}} \right)^{\frac{4n-q}{2n}},
\end{equation}
and for the potential one has
\begin{equation}\label{26}
V(\phi)  =  V_{0}\left( \phi - \phi_{0} \right)^{\frac{2nq}{q-4n}},
\end{equation}
where
\begin{widetext}
\begin{equation}\label{26.5}
V_{0} =  \dfrac{\left[  6n - (2n-1)q\right]  \rho_{\phi}^{(0)}}{6n}\left[ \dfrac{2n}{4n-q}\left( \dfrac{q\rho_{\phi}^{(0)}2^{n-1}}{3F_{0}n} \right)^{\frac{1}{2n}}\left( \dfrac{3M_{p}^{2}}{\rho_{r}^{(0)}}\right) ^{\frac{1}{2}}\right] ^{\frac{2nq}{q-4n}}.
\end{equation}
\end{widetext}
The  Einstein scalar field  equations (\ref{EqfieldSf}) (with this potential) have a special  solution. So  to study  spatially homogeneous perturbations about this special solution, we need to linearize  the scalar field equation of motion. Therefore according to
Eq.(\ref{p1(t)-diffeEq}),  the linearized equation in this epoch is given as
\begin{eqnarray}\label{39}
\dot{u_{1}}(t) & = & 0 + p_{1}(t),\\
\dot{p_{1}}(t) & = &  4\left[ {q\o 2}-{3n \o 2n-1}\right] u_1(t)+( q-5) p_1(t).\label{40}
\end{eqnarray}
 The eigenvalues of small oscillations around critical point are given by
\begin{eqnarray}\label{41}
\lambda_{n, q} = f(n, q) \pm ig(n, q),
\end{eqnarray}
where
\begin{eqnarray}\label{42}
f(n, q) & = & -\dfrac{(5-q)}{2},\nonumber \\
g(n, q) & = & \left[ \dfrac{48n-(25-2q+q^{2})(2n-1)}{4(2n-1)}\right] ^{\frac{1}{2}},
\end{eqnarray}
these eigenvalues for
$ 0 < q < 3 $
and different
$ n $,
are given at Table \ref{TabI}, (as was mentioned before, for $ 0 < q <3$ the ratio of scalar field density to radiation density, is an increasing function of time).
It is shown that(from Table \ref{TabI}) for all values of $n=1, 1.5, 2 $  and $q= 1, 2, 2.5$
 the eigenvalues $\lambda$ are $\lambda=\alpha\pm i\beta$
which $\alpha<0$ and $\beta\neq0$ , then all  critical points are spiral and stable.
\begin{table}[ht]
  \centering
  {\footnotesize
  \centering
  \begin{tabular}{p{1.1cm}||p{1.4cm}p{1.8cm}p{1.9cm}p{0.3cm}}    \\[-2mm]
                               &    \ $ $   & \ \ \ \ \ \  $\lambda$                                           \ \ \      \ \ \  &  $V(\phi)\sim$                       \ \ \ &    \\[2mm]
                         \hline \\[-3mm]
                                 &  $q=1$ \ \ \ &  $-2 \pm 2.4i$           \ \ \ &  $ (\phi -\phi_{0})^{-0.7} $     \ \ \ &     \\[2mm]
          $ n=1 $          &  $q=2$ \ \ \ &   $-1.5\pm 1.2i$         \ \ \ &  $ (\phi-\phi_{0})^{-2} $     \ \ \ &     \\[2mm]
                                &  $q=2.5$ \ \ \ &   $-1.2\pm 2.3i$         \ \ \ &  $ (\phi-\phi_{0})^{-3.3} $       \ \ \ &     \\[2mm]
    \hline \\[-3mm]
                               &  $q=1$ \ \ \ &  $-2\pm 1.7i$         \ \ \ &  $ (\phi-\phi_{0})^{-0.6} $   \ \ \ &    \\[2mm]
$n=\dfrac{3}{2}$   &  $q=2$ \ \ \ & $-1.5\pm 1.7i$              \ \ \ &  $ (\phi-\phi_{0})^{-1.5} $  \ \ \ &    \\[2mm]
                               &  $q=2.5$ \ \ \ & $-1.3\pm 1.6i$              \ \ \ & $ (\phi-\phi_{0})^{-2.1} $        \ \ \ &    \\[2mm]
    \hline \\[-3mm]
                              &     $q=1$ \ \ \ &  $-2\pm 1.4i$                          \ \ \  &  $ (\phi-\phi_{0})^{-0.6} $     \ \ \ &    \\[2mm]
         $ n= 2 $        &     $q=2$ \ \ \ &  $ -1.5\pm 1.3i$          \ \ \  &  $ (\phi-\phi_{0})^{-1.3} $        \ \ \ &    \\[2mm]
                              &     $q=2.5$ \ \ \ &  $-1.2\pm 1.2i$            \ \ \  &  $ (\phi-\phi_{0})^{-1.8} $       \ \ \ &    \\[2mm]
                        \hline \\[-2mm]
  \end{tabular}
  }
     \caption{\label{TabI}\small
The eigenvalues
$\lambda$
for different values of
$n$
and
$q$,
in radiation dominated eppoch.
}
\end{table}
\subsection{Matter dominated  epoch}
We assume the matter is dominated and therefore other components of  the Universe could be omitted. According to conservation equation  of matter, Eq.(\ref{ConservationEq}) and the equation of state for this era, one can find that $\g = 3$, therefore the energy density of matter is resulted as
\begin{equation}\label{43}
\rho_{m}= \rho_{m}^{(0)}\left( \dfrac{a_{0}}{a} \right)^{3},
\end{equation}
which $ \rho_{m}^{(0)} $ is the value of energy density of matter at $ a=a_{0} $. To investigate the behaviour of $\tilde{\rho}_{m}:=\rho_{m}/\rho_{m}^{(0)}$ versus $x$ one can refer to figure FIG.1.
\begin{figure}
\includegraphics[width=7cm]{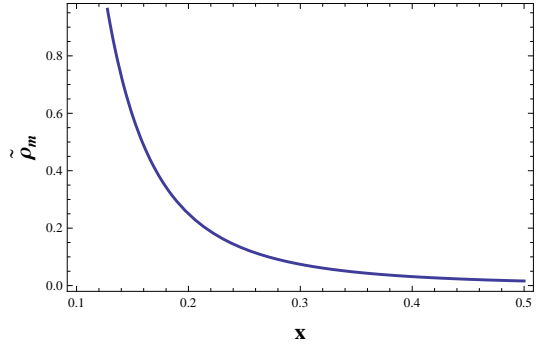}
\caption{In this figure, $\tilde{\rho}_{m}$ versus $x$ is plotted.}
\end{figure}
So one can find that
\begin{equation}\label{46}
 \phi - \phi_{0} = \dfrac{2n}{3n-q}\left[ \dfrac{q\rho_{\phi}^{(0)}2^{n-1}}{3F_{0}n} \right] ^{\frac{1}{2n}}\left( \dfrac{3M_{p}^{2}}{\rho_{m}^{(0)}}\right) ^{\frac{1}{2}}\left( \frac{a}{a_{0}} \right)^{\frac{3n-q}{2n}},
\end{equation}
and $\phi_{0}$ is the value of $\phi$ at $a=a_{0}$, and the potential comes to following form
\begin{equation}\label{49}
V(\phi)=V_{0}\left( \phi-\phi_{0}\right)^{\frac{2nq}{q-3n}},
\end{equation}
where
\begin{widetext}
\begin{equation}\label{50}
V_{0} =  \dfrac{\left[ 6n - (2n-1)q \right]  \rho_{\phi}^{(0)}}{6n}\left[ \dfrac{2n}{3n-q}\left( \dfrac{q\rho_{\phi}^{(0)}2^{n-1}}{3F_{0}n} \right)^{\frac{1}{2n}}\left( \dfrac{3M_{p}^{2}}{\rho_{m}^{(0)}}\right) ^{\frac{1}{2}} \right] ^{\frac{2nq}{q-3n}},
\end{equation}
\end{widetext}
Hence, such as radiation epoch we can find the linearized equation as
\begin{eqnarray}\label{63}
\dot{u_{1}}(t) & = & 0 + p_{1}(t),\\
\dot{p_{1}}(t) & = &  3\left[ {q\o 2}-{3n \o 2n-1}\right] u_1(t)+\left[ q-{9 \o 2}\right] p_1(t),\label{64}
\end{eqnarray}
the eigenvalues of small oscillations around critical point are given by
\begin{eqnarray}\label{65}
\lambda_{n, q} = f(n, q) \pm ig(n, q)
\end{eqnarray}
where
\begin{eqnarray}\label{66}
f(n, q) & = & -\dfrac{(9-2q)}{4}\nonumber \\
g(n, q) & = & \left[ \dfrac{144n-(81-12q+4q^{2})(2n-1)}{16(2n-1)}\right] ^{\frac{1}{2}},
\end{eqnarray}
these eigenvalues for $ 0 < q < 3 $ and different values of $n$ are brought at Table \ref{TabII}, (as was mentioned before,  for $0< q<3 $  the ratio of scalar field  to matter  energy densities, is an increasing  function of time).
It is shown that(from Table \ref{TabII}) for all values of $n=1, 1.5, 2 $  and $q= 1, 2, 2.5$ the eigenvalues $\lambda$ are $\lambda=\alpha\pm i\beta$ which $\alpha<0$ and $\beta\neq0$ , then all  critical points are spiral and stable.
\begin{table}[ht]
  \centering
  {\footnotesize
  \centering
  \begin{tabular}{p{1.1cm}||p{1.4cm}p{1.8cm}p{1.9cm}p{0.3cm}}    \\[-2mm]
                               &    \ $ $   & \ \ \ \ \ \ $\lambda$                             \ \ \     \ \ \  &  $V(\phi)\sim$                       \ \ \ &    \\[2mm]
                          \hline \\[-3mm]
                                &  $q=1$ \ \ \ &  $-1.7\pm 2.1i$                            \ \ \ &  $ (\phi-\phi_{0})^{-1} $     \ \ \ &     \\[2mm]
          $ n=1 $          &  $q=2$ \ \ \ &   $-1.2\pm 2.1i$                            \ \ \ &  $ (\phi-\phi_{0})^{-4} $     \ \ \ &     \\[2mm]
                                &  $q=2.5$ \ \ \ &   $-1\pm 2i$                                              \ \ \ &  $ (\phi-\phi_{0})^{-10} $       \ \ \ &     \\[2mm]
    \hline \\[-3mm]
                               &  $q=1$ \ \ \ &  $-1.7\pm 1.5i$         \ \ \ &  $ (\phi-\phi_{0})^{-0.9} $   \ \ \ &    \\[2mm]
$n=\dfrac{3}{2}$   &  $q=2$ \ \ \ & $-1.2\pm 1.5i$              \ \ \ &  $ (\phi-\phi_{0})^{-4} $  \ \ \ &    \\[2mm]
                               &  $q=2.5$ \ \ \ & $-1\pm 1.4i$              \ \ \ & $ (\phi-\phi_{0})^{-3.7} $        \ \ \ &    \\[2mm]
    \hline \\[-3mm]
                              &     $q=1$ \ \ \ &  $-1.7\pm 1.1i$                          \ \ \  &  $ (\phi-\phi_{0})^{-0.8} $     \ \ \ &    \\[2mm]
         $ n= 2 $        &     $q=2$ \ \ \ &  $-1.2\pm1.2i$          \ \ \  &  $ (\phi-\phi_{0})^{-2} $        \ \ \ &    \\[2mm]
                              &     $q=2.5$ \ \ \ &  $-1\pm 1.1i$           \ \ \  &  $ (\phi-\phi_{0})^{-2.8} $        \ \ \ &    \\[2mm]
                         \hline \\[-3mm]
  \end{tabular}
  }
     \caption{\label{TabII}\small
The eigenvalues
$\lambda$
for different values of
$n$
and
$q$,
in matter dominated epoch.
}
\end{table}

\subsection{Scalar  field dominated  epoch}
We assume the scalar field energy density is dominated and therefore we neglect the other components of the Universe. According to conservation equation for scalar field, Eq.(\ref{ConservationEq}) and the equation of state of scalar field, $p_{\phi}=(q/3 -1)\rho_{\phi}$, one finds $\g =q$, therefore the scalar field  energy density is obtained as
\begin{equation}\label{66.5}
\rho_{\phi} = \rho_{\phi}^{(0)}\left( \dfrac{a_{0}}{a} \right)^{q},
\end{equation}
which $ \rho_{\phi}^{(0)} $, is the value of scalar field energy density at $ a=a_{0} $. For more details one can refer to figure FIG.2, where shows the behaviour of $\tilde{\rho}_{\phi}:=\rho_{\phi}/\rho_{\phi}^{(0)}$ versus $x$.
\begin{figure}
\includegraphics[width=7cm]{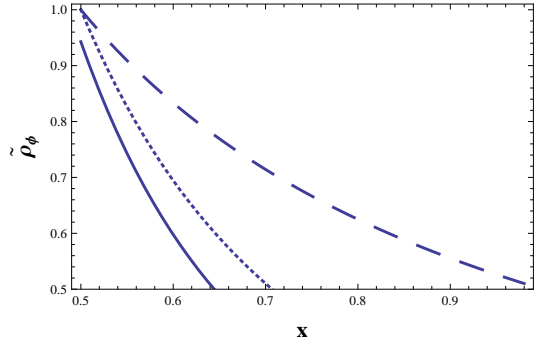}
\caption{In this figure, $\tilde{\rho}_{\phi}$ versus $x$ is plotted. Where the dashed, dotted and solid lines are devoted to $q=1, 2, 2.5$ respectively.}
\end{figure}
 So by substituting  $q$ instead of $\g$ in Eqs. (\ref{FreEqGamma}) and integrating it  for $n=1$,   we have
\begin{equation}\label{67}
\phi - \phi_{0}=\sqrt{ qM_P^2}\ln({a\o a_0}),
\end{equation}
where $\phi_{0}$ is the value of $\phi$ at $a=a_{0}$. So using Eqs. (\ref{VPhi2}) and (\ref{67}) we have
\begin{equation}\label{71}
V(\phi)=\left[{6-q \o 6}\right]\rho_{\phi}^{(0)} e^{-{\sqrt{q}\o M_P}(\phi-\phi_0)},
\end{equation}
and for $n \neq 1$ we obtain
\begin{equation}\label{69}
\phi - \phi_{0} = \dfrac{2n}{q(n-1)}\left[ \dfrac{q\rho_{\phi}^{(0)}}{6F_{0}n}\right] ^{\frac{1}{2n}}\left( \dfrac{6M_{p}^{2}}{\rho_{\phi}^{(0)}}\right) ^{\frac{1}{2}}\left( \frac{a}{a_{0}} \right)^{\frac{q(n-1)}{2n}},
\end{equation}
 and
 \begin{equation}\label{71}
V(\phi)=V_{0}(\phi-\phi_{0})^{\frac{2n}{1-n}},
\end{equation}
where
\begin{widetext}
\begin{equation}\label{72}
V_{0} =  \dfrac{\left[ 6n - (2n-1)q \right] \rho_{\phi}^{(0)}}{6n}\left[  \dfrac{2n}{q(n-1)}\left( \dfrac{q\rho_{\phi}^{(0)}}{6F_{0}n} \right)^{\frac{1}{2n}}\left( \dfrac{6M_{p}^{2}}{\rho_{\phi}^{(0)}}\right) ^{\frac{1}{2}}\right] ^{\frac{2n}{n-1}}.
\end{equation}
\end{widetext}
Also the linearized form of  Einstein scalar field  equation (with these potentials)  is given by
\begin{eqnarray}\label{85}
\dot{u_{1}}(t) & = & 0 + p_{1}(t),\\
\dot{p_{1}}(t) & = &  q\left[ {q\o 2}-{3n \o 2n-1}\right] u_1(t)+\left[ {q \o 2}-3\right] p_1(t),\label{85.5}
\end{eqnarray}
the eigenvalues of small oscillations around critical point are given by
\begin{eqnarray}\label{86}
\lambda_{n, q} = f(n, q) \pm ig(n, q)
\end{eqnarray}
where
\begin{eqnarray}\label{87}
f(n, q) & = & -\dfrac{8-q}{4}\nonumber \\
g(n, q) & = & \left[ \dfrac{80nq-[9q^{2}+64][2n-1]}{16(2n-1)}\right] ^{\frac{1}{2}},
\end{eqnarray}
these eigenvalues for
$ 0 < q < 3 $
and different
$ n $,
are given at Table \ref{TabIII}.
\begin{table}[ht]
  \centering
  {\footnotesize
  \centering
  \begin{tabular}{p{1.1cm}||p{1.4cm}p{1.8cm}p{1.9cm}p{0.3cm}}    \\ [-2mm]
                               &    \ $ $   &  \ \ \ \ \ \  $\lambda$                                \ \ \    \ \ \  &  $V(\phi)\sim$                    \ \ \ &    \\[2mm]
                       \hline \\[-3mm]
                                &     $q=1$ \ \ \ &  $-1.7\pm 0.7i$                             \ \ \ &  $e^{-{(\phi-\phi_0)\o M_P}} $ \ \ \ &     \\[2mm]
          $ n=1 $          &     $q=2$ \ \ \ &   $-1.5\pm 1.9$                              \ \ \ &  $e^{-{\sqrt{2}(\phi-\phi_0)\o M_P}} $     \ \ \ &     \\[2mm]
                                &     $q=2.5$ \ \ \ &   $-1.4\pm 2.2i$              \ \ \ &  $ e^{-{\sqrt{2.5}(\phi-\phi_0)\o M_P}} $     \ \ \ &    \\[2mm]
    \hline \\[-3mm]
                               &  $q=1$ \ \ \ &  $-1.7\pm 0.9$      \ \ \ &  $ (\phi-\phi_{0})^{-6} $ \ \ \ &       \\[2mm]
$n=\dfrac{3}{2}$   &  $q=2$ \ \ \ &  $-1.5\pm 1.1i$        \ \ \ &  $ (\phi-\phi_{0})^{-6} $ \ \ \ &       \\[2mm]
                               &  $q=2.5$ \ \ \ &  $-1.4\pm 1.3i$          \ \ \ & $ (\phi-\phi_{0})^{-6} $     \ \ \ &       \\[2mm]
    \hline \\[-3mm]
                              &     $q=1$ \ \ \ &  $-1.7\pm 1.1$                      \ \ \  &  $ (\phi-\phi_{0})^{-4} $   \ \ \ &    \\[2mm]
         $ n= 2 $        &     $q=2$ \ \ \ &  $-1.5\pm 0.7i$             \ \ \  &  $ (\phi-\phi_{0})^{-4} $   \ \ \ &     \\[2mm]
                              &     $q=2.5$ \ \ \ &  $-1.4\pm 0.9i$                   \ \ \  &  $ (\phi-\phi_{0})^{-4} $ \ \ \ &    \\[2mm]
                        \hline \\[-3mm]
  \end{tabular}
  }
     \caption{\label{TabIII}\small
The eigenvalues
$\lambda$
for different values of
$n$
and
$q$,
in scalar field dominated eppoch.
}
\end{table}
It is clearly seen that all eigenvalues which are complex such as $\lambda=\alpha\pm \beta i$ which $\alpha<0$ and $\beta\neq 0$, describe critical points which are  spiral critical point and stable. For $n=1$ and $q=2$ the eigenvalues are $\lambda_1 =0.4 $ and $\lambda_2=-3.4 $ so the critical point is a saddle point. For $n=3/2, 2$ and $q=1$ the different eigenvalues are negative and stable.\\
Now we want to estimate the mass of the scalar field, $m_\phi$, in the scalar field dominated era. It is obvious that by expanding the scalar field around the background field, one can suppose, $m_\phi^2=d^2V(\phi)/2d\phi^2\mid_{\phi=\phi_0}.$ Hence for the potential function which was obtained for $n\neq1$ in (\ref{71}), one has
\begin{eqnarray}\label{88}
m_\phi^2=\frac{d^2V(\phi)}{2d\phi^2}=nV_0\frac{3n-1}{(1-n)^2}\Big(\phi-\phi_0\Big)^{\frac{5n-3}{1-n}}.
\end{eqnarray}
By substituting Eqs.(\ref{69}) and (\ref{72}) in above equation, the mass of scalar field could be obtained as
$m_\phi=3.5\times10^{-25} \Big(\frac{a_{0}}{a_{t}}\Big)^{0.1} cm^{-1}.$
To get this value, we have taken $n = 1.2,  q = 2.5, F_{0} = 7.0,  \rho_{\phi}^{(0)}= 3.5 \times10^{-25} g/cm^3$  and  $M_{p}= 2.5\times10^{-5} g.$
It should be noted this value is in good agrement with previous works {\cite{5a}}.
\section{conclusion and discussion}
It is undoubtedly accepted, the scalar field mechanism attracts more attention in recent semi-century.  For instance a wide range of problems and concepts  such as hierarchy problem, steady state cosmology models, inflation,  CDM, dark energy and so on by means of scalar field mechanism have been investigated. Amongst all different proposals of scalar field mechanism we have considered the non-canonical scenario, because of it's advantages in solving coincidence and fine tuning problems. For such model we successfully have obtained a definite form of scalar field potential. It was shown that for definite scalar field potential, in any related epoch, the scalar field solutions for an appropriate range of parameters are stable. In fact, it has indicated that, for example in radiation and matter dominant eras, all eigenvalues of small oscillations around critical point are as $\lambda = \alpha+i\beta$, where $\alpha < 0$ and $\beta\neq 0 $. Also it should be noted all critical points were spiral and stable. In addition, in scalar field epoch for $(n, q)= (1, 2)$ the eigenvalues are $\lambda_1 = 0.4, \lambda_2 = -3.4 $, hence the critical point is a saddle point, for $(n, q)= (3/2, 1)$ and $(2, 1)$, the different eigenvalues are negative, therefore the critical point is stable. Also for scalar field dominated era, by means of the some well-known values for this epoch the mass of scalar field has been obtained as $m_\phi=3.5\times10^{-25} (a_{0}/a_{t})^{0.1} cm^{-1}$.
\\
From the results which are shown in tables it is seen that for $ n\neq 1 $ in all epoch the potential has inverse power law form, and this indicates that the better candidate potential for explaining the evaluation of universe especially in inflation epoch has inverse power law form. Furthermore, our results show that for different form of noncanonical model we have different potential but all of  them are inverse power law.
\section{Acknowledgment}
H. Sheikhahmadi would like to thank Iran's National Elites Foundation for financially support during this work.


\begin{thebibliography}{99}
\bibitem{CMB_th_1} G. Efstathiou  and J. R. Bond,  Mon. Not. Roy. Astron. Soc  \textbf{304},  75 (1999).
\bibitem{CMB2}  J. Dunkley, et al.  Astrophys. J  \textbf{701},  1804 (2009).
\bibitem{ch1:7}A. G. Riess  et al.,  Astron. J \textbf{116},  1009  (1998).
\bibitem{ch1:8} S. Perlmutter,  et al., Astrophys. J \textbf{517},  565 (1999).
\bibitem{BAO_th_Eisenstein} D. J. Eisenstein and  W. Hu,  Astrophys. J. \textbf{496},  605 (1998) .
\bibitem{BAO_th_Shoji} M. Shoji, D. Jeong and   E. Komatsu,  Astrophys. J.  \textbf{693},  1404  (2009).
\bibitem{FarooqRatra} O. Farooq and B. Ratra, ,  arXiv: 1301.5243 (2013).
\bibitem{H0_1}  G. Chen, J. R. Gott and B. Ratra,    Publ. Astron. Soc. Pac.  \textbf{115},  1269 (2003).
\bibitem{ch1:10} M. Tegmark,   et al., Phys. Rev. D \textbf{69},  103501 (2004).
\bibitem{SD2} J. Sollerman, et al.,  Astrophys. J. \textbf{703}, 1374 (2009) .
 \bibitem{ch1:9} D. N. Spergel, et al.,  Astrophys. J. suppl \textbf{148},  175 (2003).
\bibitem{WM2} G. Hinshaw, et al.,   Astrophys. J. Suppl  \textbf{180}, 225 (2009) .
\bibitem{Bamba} K. Bamba,  S. Capozziello, S. Nojiri,  and   S. D. Odintsov, arXiv:gr-qc/1205.3421 (2012).
\bibitem{Ade01} P. A. R. Ade et al.,  A \& A \textbf{571}, A11 (2014).
\bibitem{Ade02} Ade, P. A. R.  et al. Phys. Rev. Lett. \textbf{112}, 241101 (2014).
\bibitem {3} A. Einstein, Sitzungsber. K. preuss. Akda. Wiss. {\bf 142} (1917), [The prnciple
of relativity (Dover, New York, 1952), P. 177].
\bibitem {3a} S. Weinberg,  Rev. Mod. Phys. {\bf 61}, 1 (1989).
\bibitem {3d} V.  Sahni, A. A. Starobinisky,  Int. J .Mod. Phys. D {\bf 9}, 373 (2000).
\bibitem{3b} P. J. E.  Peebles,  B. Ratra,  Rev. Mod. Phys. {\bf 75}, 559 (2003).
\bibitem{3e} T. Padmanabha, Phys. Rep. {\bf 380}, 235 (2003).
\bibitem {4} Steinhardt P. J, in critical problems in physics, edited by V. L. Fitch and D. R. Marlow
(Printed University Press, Prinston, NJ), (1997).
\bibitem {fr1}  D. Wands, Class. Quant. Grav. \textbf{11}, 269 (1994).
\bibitem {fr2}  S. Nojiri,  and  S. D. Odintsov,  J. Phys. A \textbf{40}, 6725 (2007).
\bibitem {fr3} A. Aghamohammadi, Kh. Saaidi,  M. R. Abolhassani,  A. Vajdi,   Int. J. Theor. Phys. \textbf{49} 709 (2010).
\bibitem {fr4} Kh. Saaidi,   A. Vajdi, S. W. Rabiei, A.  Aghamohammadi, H. Sheikhahmadi,  Astro Spac Scie \textbf{337}, 739, (2012).
\bibitem {ft1} R. Ferraro, F. Fiorini, Phys. Rev. D \textbf{75}, 084031 (2007).
\bibitem {ft2} R. Ferraro, F. Fiorini, Phys. Rev. D \textbf{ 78}, 124019 (2008).
\bibitem {ft3} K. Karami, A. Abdolmaleki, JCAP \textbf{04}, 007 (2012).
\bibitem {ft4} A. Aghamohammadi, Astrophys. Space. Sci. \textbf{352}, 1  (2014).
\bibitem {ft5} H. Sheikhahmadi, A. Aghamohammadi, Kh. Saaidi, Phys. Lett. B \textbf{749}, 231 (2015).
\bibitem {fg1} S. Capozziello, V. F. Cardone and V. Salzano, Phys. Rev. D \textbf{78}, 063504 (2008).
\bibitem {fg2} S. Capozziello, V. F. Cardone, arXiv:0902.0088 [astro-ph], (2009).
\bibitem {fg3} M. Bouhmadi - Lopez, S. Capozziello, V. F. Cardone, Phys. Rev. D \textbf{82}, 103526 (2010).
\bibitem {fg4} S. Capozziello1, V. F. Cardone, H. Farajollahi, A. Ravanpak, Phys. Rev. D \textbf{84}, 043527 (2011).
\bibitem {13O} A. Cohen, D. Kaplan, and A. Nelson, Phys. Rev. Lett. \textbf{82}, 4971 (1999).
\bibitem {13aO} B. Guberina, R. Horvat, and H. Nikolic, JCAP\textbf{ 0701},012 (2007).
\bibitem {13bO} L. Susskind, J. Math. Phys. (NY)\textbf{ 36}, 6377 (1995).
\bibitem {13cO} M. Li, Phys. Lett. B \textbf{603} (2004).
\bibitem {13dO}  K. Enqvist and M. S. Sloth, Phys. Rev. Lett. \textbf{93}, 221302 (2004).
\bibitem {14O} M. Maziashvili,  Int. J. Mod. Phys. D {\bf 16}, 1531 (2007).
\bibitem {14aO} M. Maziashvili, Phys. Lett. B {\bf 652}, 165  (2007).
\bibitem {15O} H. Wei and R. G. Cai, Phys. Lett. B {\bf 660}, 113 (2008).
\bibitem {12O}  L. Hollenstein, M. Jaccard, M. Maggiore, and E. Mitsou, Phys. Rev. D \textbf{85}, 124031 (2012).
\bibitem {11O} M. Maggiore, Phys. Rev. D \textbf{83}, 063514 (2011).
\bibitem {11aO}  S. A. Fulling, L. Parker, Annals Phys \textbf{87},  176 (1974).
\bibitem {My} H. Sheikhahmadi, A. Aghamohammadi and Kh. saaidi, arXiv:1407.0125[gr-qc] (2014).
\bibitem {Yu} Y. B. Zeldovich and M. Y. Khlopov,  Phys. Lett. B\textbf{79},  239 (1978).
\bibitem {Yu1} J. Preskill, Phys. Rev. Lett. \textbf{43},  1365 (1979).
\bibitem {L-G} V. L. Ginzburg and L. D. Landau,  Zh. Eksp. Teor. Fiz. \textbf{20},  1064 (1950).
\bibitem {H} P. W. Higgs,  Phys. Rev. Lett. 13,  508 (1964).
\bibitem {H1} F. Englert and R. Brout,  Phys. Rev. Lett. 13,  321 (1964).
\bibitem {H2} G. S. Guralnik, C. R. Hagen, and T. W. B. Kibble, Phys. Rev. Lett. 13, 585 (1964).
\bibitem {I} A. H. Guth, Phys. Rev. D \textbf{23},  347 (1981).
\bibitem {I1} A. D. Linde, Phys. Lett. B \textbf{108},  389 (1982).
\bibitem {I2} A. J. Albrecht and P. J. Steinhardt, Phys. Rev. Lett. \textbf{48},  1220 (1982).
\bibitem {Ia} T.  Golanbari, A. Mohammadi,  and Kh. Saaidi, Phys. Rev. D. \textbf{89}, 103529 (2014).
\bibitem {Ia1} A. Aghamohammadi, A. Mohammadi, T. Golanbari, Kh. Saaidi, Phys. Rev. D. \textbf{90}, 084028 (2014).
\bibitem {DBI} Alishahiha, E. Silverstein, and D. Tong, Phys. Rev. D \textbf{70 }, 123505 (2004).
\bibitem {brans1} C. H. Brans and R. H. Dicke, Phys. Rev. \textbf{124}, 925 (1961).
\bibitem {brans2} C. H. Brans and R. H. Dicke, Phys. Rev. \textbf{125}, 2194 (1962).
\bibitem {5}  P. J. E.  Peebles,  B. Ratra, Astrophys. J. Lett. {\bf 17}, 325 (1988).
\bibitem {5a} B. Ratra, P. J. E. Peebles,  Phys. Rev. D {\bf 37}, 3406 (1988).
\bibitem {5c} C. Wetterich,   Nucl. Phys. B {\bf 302}, 668 (1988).
\bibitem {5b} T. G. Clemson, A. R. Liddle , Mon. Not.Roy. Astron. Soc. {\bf 395}, 1585 (2009).
\bibitem {6} C.  Armendariz-Picon,  V. F. Mukhanov,  P. J. Steinhardt,  Phys. Rev.Lett. {\bf 85}, 4438 (2000).
\bibitem {6a} T. Chiba , T. Okabe , M. Yamaguchi , Phys. Rev. D \textbf{62}, 023511 (2000).
\bibitem {6b} Armendariz-Picon, C.  Mukhanov, V. F.  and Steinhardt, P. J. Phys.Rev. D {\bf 63}, 103510 (2001).
\bibitem {7} A. Sen,  JHEP {\bf 0207}, 065 (2002).
\bibitem {8} R. R. Caldwell, Phys. Lett. B {\bf 545}, 23 (2002).
\bibitem {8-1} R. R. Caldwell,  M. Kamionkowski, N. N. Weinberg,  Phys. Rev. Lett. {\bf 91}, 071301 (2003).
\bibitem {8a} X. Chen, Y. Gong, E. N. Saridakis, {JCAP} \textbf{0904}, 2009 001.
\bibitem {8b} J. M.  Cline, S. Y.  Jeon,  G. D. Moore, Phys. Rev. D {\bf 70}, 043543 (2004).
\bibitem {9} E. Elizalde,   S.  Nojiri, S. D. Odinstov,   Phys. Rev. D {\bf 70}, 043539 (2004).
\bibitem {9a} S. Nojiri, S. D.  Odintsov, S.  Tsujikawa,  Phys. Rev. D {\bf 71}, 063004 (2005).
\bibitem {9b} A. Anisimov,   E. Babichev, A.  Vikman,  J. Cosmol. Astropart. Phys. {\bf 06}, 006 (2005).
\bibitem {11} J. Khoury, A.  Weltman,  Phys. Rev. Lett. {\bf 93}, 171104 (2004).
\bibitem {11a} J. Khoury,   A. Weltman,   Phys. Rev. D {\bf 69}, 044026 (2004).
\bibitem {10} D. F. Mota,  J. D. Barrow,   Phys. Lett. B {\bf 581}, 141 (2004).
\bibitem{ch18} Kh. Saaid, H. Sheikahmadi and J. Afzali,  Astrophys. Space. Sci. \textbf{333}  501 (2011).
\bibitem{ch19} Kh. Saaidi, A. Mohammadi, H. Sheikhahmadi,  Phys. Rev. D. \textbf{83} 104019 (2011).
\bibitem{ch20} Kh. Saaidi, H. Sheikhahmadi, T. Golanbari,   and S.W. Rabiei,  Astrophys. Space. Sci. \textbf{348}   233 (2013).
\bibitem{ch20-1}T. Golanbari, A. Mohammadi, Kh. Saaidi,  Int. J. Mod. Phys. A \textbf{29}, 1450033 (2014).
\bibitem{ch1:20a} A. Aghamohammadi, Kh. Saaidi, A. Mohammadi, H. Sheikhahmadi, T. Golanbari,   and S.W. Rabiei, Astrophys. Space. Sci.  {\bf 345}, 17 (2013).
 \bibitem{ch1:18} S. Carroll, Phys. Rev. Lett. \textbf{81} 3067 ( 1998).
\end{thebibliography}
\end{document}